\documentclass[aps,prb,twocolumn,superscriptaddress,10pt]{revtex4-2}
\setcitestyle{numbers,square}
\pdfoutput=1
\usepackage{amsmath,amssymb,amsfonts,upgreek}
\usepackage{color}
\usepackage{graphicx}
\usepackage{siunitx}
\newcommand{\siev}[1]{\SI{#1}{\eV}}
\DeclareSIUnit\elc{\mbox{$e^-$}}
\DeclareSIUnit\ry{\mbox{Ry}}
\usepackage[nolist,nohyperlinks]{acronym}

\usepackage{hyperref}
\hypersetup{colorlinks=true,linkcolor=red,citecolor=red}
\usepackage{cleveref}
\Crefname{figure}{Fig.}{Figs.}

\newcommand{\svo}{SrVO$_3$}
\newcommand{\lvo}{LaVO$_3$}
\newcommand{\sto}{SrTiO$_3$}
\newcommand{\lao}{LaAlO$_3$}
\newcommand{\ttg}{$t_{2g}$}
\newcommand{\eg}{$e_g$}
\newcommand{\umit}{\ensuremath{U_{\mathrm{MIT}}}}
\newcommand{\dftu}{\ac{DFT}\ensuremath{+U}}
\newcommand{\done}{\ensuremath{d^1}}
\newcommand{\dtwo}{\ensuremath{d^2}}

\begin{document}

\title{Interfacial doping in \lvo{}/\svo{} multilayers from DFT+DMFT}

\author{Sophie Beck}
\email{sbeck@flatironinstitute.org}
\affiliation{Materials Theory, ETH Z\"u{}rich, Wolfgang-Pauli-Strasse 27, 8093 Z\"u{}rich, Switzerland}
\affiliation{Center for Computational Quantum Physics, Flatiron Institute, 162 5th Avenue, New York, NY 10010, USA}
\author{Claude Ederer}
\email{claude.ederer@mat.ethz.ch}
\affiliation{Materials Theory, ETH Z\"u{}rich, Wolfgang-Pauli-Strasse 27, 8093 Z\"u{}rich, Switzerland}

\date{\today}

\begin{abstract}

We investigate the effect of spatial doping of the Mott insulator \lvo{} by inserting a few layers of the correlated metal \svo{} in multilayer geometries.
Using \acl{DFT} in combination with \acl{DMFT}, we demonstrate that this leads to a geometrically confined and robust metallic layer that stabilizes the metallicity in \svo{} even in the ultrathin layer limit, suppressing a potential dimensionality-induced metal-insulator transition.
For a thicker \svo{} layer, we find a continuous transition of both structural and electronic properties across the interface between the two materials, with bulk properties reestablished on a length scale of \numrange{2}{3} unit cells away from the interface. 
We show that a strain modulation applied along the growth direction can lead to asymmetric charge reconstruction at chemically symmetric interfaces.
However, we find that this effect is rather weak, implying that fractional occupancy, and thus metallicity, persists at the interfaces.

\end{abstract}

\maketitle

\begin{acronym}
 \acro{DFT}{density functional theory}
 \acro{DMFT}{dynamical mean-field theory}
 \acro{MIT}{metal-insulator transition}
\end{acronym}
\acresetall

\section{Introduction}

Heterostructures constructed from complex oxides show great potential for novel electronic applications due to a variety of possible interfacial reconstruction effects~\cite{Mannhart/Schlom:2010,Zubko_et_al:2011,Hwang_et_al:2012,Chakhalian/Millis/Rondinelli:2012}.
Many of these systems show properties fundamentally different from those of the corresponding bulk compounds. Prominent examples are the quasi-two-dimensional electron gas and the interfacial magnetism emerging in between the otherwise non-magnetic band insulators \lao{} and \sto{}~\cite{Ohtomo/Hwang:2004,Brinkman_et_al:2007}.

The phenomenon of emerging phases is also present in simple solid solutions, where the gradual increase in cation substitution often leads to a transition region exhibiting properties that differ fundamentally from those of the two end compounds.
One example is La$_{1-x}$Sr$_x$VO$_3$, where gradual changes of structural, spin, charge, and orbital degrees of freedom as a function of composition result in a transition to an antiferromagnetic metallic phase for $0.178 \leq x \leq 0.260$~\cite{Miyasaka/Okuda/Tokura:2000} that is not present in either of the constituent compounds, with \lvo{} being an antiferromagnetic Mott insulator and \svo{} a correlated metal.

On the other hand, whereas emerging properties in solid solutions are solely governed by the overall composition, artificial superlattices offer the spatial distribution as an additional powerful tuning parameter, allowing, e.g., for geometrically confined doping or variations in the individual layer thicknesses and periodicity.
For example, an experimental study on (\lvo{})$_{m}$/(\svo{})$_1$ superlattices reported a filling-controlled insulator-to-metal transition upon reducing \mbox{$m \leq 4$}, corresponding to a doping concentration of \mbox{$x = 0.2$}~\cite{Sheets/Mercey/Prellier:2007}.
However, a follow-up study then reported consistently metallic behavior for samples from $m=2$ to $m=6$ based on in-plane transport measurements, and it was suggested that pressure during sample preparation may play an important role~\cite{Luders_et_al:2009}. 
Interestingly, the same multilayers exhibited an insulating out-of-plane resistance, in contrast to the in-plane measurements, demonstrating the more complex properties emerging in superlattices compared to simple solid solutions.

Furthermore, the \svo{} layers buried in \lvo{} do not show a dimensionality-induced \ac{MIT}, as reported, e.g.,  for ultra-thin films of \svo{} grown on \sto{}~\cite{Yoshimatsu_et_al:2010,Zhong_et_al:2015} and also for many other correlated metals. Such a transition generally results from the reduced bandwidth and orbital splittings due to symmetry lowering~\cite{Scherwitzl_et_al:2011, Gu_et_al:2013, Biswas/Kim/Jeong:2014, Liao_et_al:2015,Zhong_et_al:2015, Beck_et_al:2018}.
Instead, the dimensional crossover of an ultra-thin \svo{} film (3 unit cells) buried within \lvo{} layers results in a weakly localized two-dimensional electron liquid state~\cite{Li_et_al:2015}, promoting the idea of \svo{} acting as a geometrically confined dopant within \lvo{}~\cite{Luders_et_al:2009}.

This picture is also supported by optical spectroscopy on (\lvo{})$_{6n}$/(\svo{})$_n$ superlattices~\cite{Jeong_et_al:2011}, for which valence changes differing from the nominal $3d^2$ and $3d^1$ valence in \lvo{} and \svo{}, respectively, were observed using STEM-EELS measurements~\cite{Boullay_et_al:2011}.
Although details could not be inferred at that time due to limited spatial resolution, a follow-up study investigated the valence states in (\lvo{})$_6$/(\svo{})$_3$ superlattices more closely and with atomic resolution, showing an asymmetric charge distribution at chemically symmetric interfaces, which was suggested to result from an asymmetry in the out-of-plane lattice parameter due to strain relaxation~\cite{Tan_et_al:2013}.
The same study further revealed clear signatures of the mixed valence in the superlattice and a length scale for electronic reconstruction of \SIrange{0.5}{1.2}{\nm}, in accordance with previous studies on related complex  oxides~\cite{Okamoto/Millis:2004,Cantoni_et_al:2012,Hwang_et_al:2012}.

A theoretical study on (\lvo{})$_{m}$/(\svo{})$_1$ with \mbox{$m=5,6$} used \dftu{} to show that the interfacial vanadium sites can also arrange in a checkerboard pattern of V$^{3+}$ and V$^{4+}$~\cite{Dai_et_al:2018}.
While the theoretically predicted insulating behavior in these superlattices agrees with the earlier reports in Ref.~\cite{Sheets/Mercey/Prellier:2007} and also with results from optical measurements~\cite{Jeong_et_al:2011}, a metallic character was observed in the resistivity measurements of Ref.~\cite{Luders_et_al:2009}, raising the issue of oxygen vacancies~\cite{Dai/Eckern/Schwingenschloegl:2018}. 
A charge ordered checkerboard arrangement of V$^{3+}$ and V$^{4+}$ sites has also been reported in an earlier \dftu{} study of short periodicity (\lvo{})$_1$/(\svo{})$_1$ multilayers~\cite{Park/Kumar/Rabe:2017}.

From this brief summary, it is clear that the \lvo{}/\svo{} system has attracted considerable attention due to the interesting range of physical properties of the two bulk compounds, which is further extended in artificial heterostructures with additional control parameters.
However, a conclusive picture of the electronic structure at the interface has not yet emerged, and thus first-principles calculations are required to elucidate the mechanisms of interfacial reconstruction at the atomic level.
Here, we investigate multilayers with varying thicknesses of the Mott insulator \lvo{} and the correlated metal \svo{} with regard to structural and electronic reconstruction using a combination of \ac{DFT} and \ac{DMFT}.
We show that at the \ac{DFT} level there is a gradual transition between the bulk crystal structures and that this trend is robust with variation of the multilayer thickness.
We demonstrate that charge is indeed transferred from \lvo{} to \svo{}, leaving the interfacial V layer in a ${+3.5}$ oxidation state, consistent with its mixed chemical environment, and with occupations decaying to nominal values on a length scale of \SIrange{2}{3} unit~cells at the \ac{DFT} level. 
The role of \svo{} as a dopant is discussed at the \ac{DMFT} level, where we find persistent metallicity in the interfacial layer, even if increased correlations drive \svo{} into an insulating state.
Finally, we also demonstrate that asymmetric strain relaxation at the interface can indeed induce a weak electronic asymmetry at chemically symmetric interfaces. 

The paper is structured as follows:
In \cref{sec:cm} we introduce the multilayer structures and the computational methods used in this work.
Structural and electronic results comparing the different multilayers are discussed in \cref{sec:results}.
A summary is presented in \cref{sec:sum}.

\section{Computational method}
\label{sec:cm}

\subsection{Supercell construction}
\label{subsec:construct}

To analyze the interplay between interfacial doping and the film thickness of the materials, we consider multiple superlattices, where we vary the number of layers of each component.
We consider periodic (\lvo{})$_i$/(\svo{})$_j$ multilayers, $i/j$ in shorthand notation, where $i$ and $j$ refer to the number of LaO and SrO sublayers, respectively.
The layers are stacked along the $[001]$ direction ($z$-axis of the superlattices), which is chosen parallel to the long orthorhombic axis of the bulk \lvo{} $Pbnm$ unit cell.
We note that in the literature the alternative $[110]$ $Pbnm$ stacking direction is also commonly used~\cite{Luders_et_al:2014, Sclauzero/Ederer:2015}, which is briefly discussed in Sec.~\ref{subsec:structure}.

The in-plane lattice parameters of the superlattices are fixed to the average of the pseudocubic lattice constants \mbox{$a_{\textrm{cub}} = (a/\sqrt{2} + b/\sqrt{2} + c/2)/3$} calculated for the two bulk materials; \SI{3.894}{\angstrom} for \lvo{} and \SI{3.859}{\angstrom} for \svo{}, resulting in an average value of \SI{3.877}{\angstrom}. This corresponds to a moderate lattice mismatch of \SI{\pm 0.45}{\%}, i.e., a small compressive and tensile strain for \lvo{} and \svo{}, respectively.
Thus, compared to experiment, the lattice mismatch is underestimated approximately by a factor of $2$.
While this could have a small quantitative effect on our results, we do not expect the qualitative picture to change.
We further note that the specific strain levels in the experimental samples also depend on whether the superlattices are constrained to the substrate lattice constant or are (partially) relaxed.

All multilayers are constructed with $\sqrt{2} \times \sqrt{2}$ in-plane lattice vectors relative to the pseudocubic units to accommodate the tilts and rotations of the oxygen octahedra allowed within $Pbnm$ symmetry.
We adopt the notation used in Refs.~\cite{Rondinelli/Spaldin:2011,Beck_et_al:2018} to quantify the octahedral tilt and rotation angles $\phi$ and $\theta$ (relative to the long orthorhombic $c$ axis) in terms of specific bond angles depicted in \Cref{fig:ang_LS}.
The tilt pattern in bulk \lvo{} is $a^-a^-c^+$ in Glazer notation, whereas bulk \svo{} crystallizes in cubic $Pm\bar{3}m$ symmetry without octahedral tilts and rotations ($a^0a^0a^0$).
In all multilayers presented here, a mirror plane parallel to the $z$-direction of the superlattices is preserved, which keeps $B$ sites within the same $xy$-plane symmetry-equivalent.
Since all $B$ sites throughout the multilayer are occupied by the same atomic species, additional mirror planes perpendicular to the stacking direction are maintained in the central $A$O layers if both $i$ and $j$ are chosen to be odd numbers.
This setup is used in all cases unless specified otherwise.

\subsection{\ac{DFT} + \ac{DMFT} method}

\paragraph{\ac{DFT} calculation}

We use the \textsc{Quantum~ESPRESSO} package~\cite{Giannozzi_et_al:2009} for the \ac{DFT} calculations with the generalized gradient approximation of the exchange-correlation functional in the formulation of Perdew, Burke, and Ernzerhof (PBE)~\cite{Perdew/Burke/Ernzerhof:1996} and scalar-relativistic ultrasoft pseudopotentials.
The $3s$ and $3p$ semicore states of V, the $4s$ and $4p$ semicore states of Sr, and the $5s$ and $5p$ semicore states of La, are treated as valence electrons, while the empty La-$4f$ states are not included.
The plane-wave energy cutoffs are set to \SI{70}{\ry} for the wavefunctions and to \SI{840}{\ry} for the charge density.
A Methfessel-Paxton scheme is used for the Brillouin-zone integration with a smearing parameter of \SI{0.02}{\ry}.
For the $k$-point sampling we use a $6 \times 6 \times 4$ Monkhorst-Pack grid for the smallest unit cell with $i=j=1$ and a reduced number of points along $k_z$ for all larger supercells. 
The atomic coordinates are relaxed until all force components are smaller than \SI[per-mode=symbol]{1}{\milli\ry\per\bohr} ($a_0$: Bohr radius) and the energy difference in successive steps is below \SI[per-mode=symbol]{0.1}{\ry\per\bohr}.
The $z$ component of the simulation cell is relaxed until the stress tensor component drops below \SI{0.1}{kbar}, while both in-plane lattice parameters are kept fixed to the value given above.
The effect of strain is addressed by adjusting the in-plane cell parameters accordingly and again relaxing the $z$-component.

\paragraph{\ac{DMFT} calculation}

For the \ac{DMFT} calculations we use a low-energy effective correlated subspace corresponding to the V-\ttg{}-dominated bands.
As in bulk \lvo{} and \svo{}, the corresponding Kohn-Sham bands are well separated from lower O-$p$ dominated bands and have only minimal overlap with states higher in energy, here mostly V-\eg{} dominated bands.
The downfolding is performed using the \textsc{Wannier90} code~\cite{Mostofi_et_al:2008}, resulting in a set of three \ttg{}-like maximally localized Wannier functions per V site.
We use TRIQS/solid\_dmft~\cite{Merkel_et_al:2022} based on the TRIQS/DFTTools library~\cite{aichhorn_dfttools_2016} to run the \ac{DMFT} cycle, where a self-consistent solution to the lattice problem is found when the local lattice Green's function matches that of the effective impurity problem.
The local electron-electron interaction is parametrized in the Slater-Kanamori form, in terms of the inter-orbital Hubbard $U$ and the Hund coupling $J$, including spin-flip and pair-hopping terms~\cite{Castellani/Natoli/Ranninger:1978}.
A double counting correction in the fully localized limit is subtracted~\cite{Solovyev:1994,anisimov1997}.
To solve the effective impurity problem, we use a continuous-time quantum Monte Carlo hybridization-expansion solver implemented in TRIQS/CTHYB~\cite{Seth2016274} with inverse temperature $\beta=1/(k_{\text{B}}T) = 40$ eV$^{-1}$.
To monitor the effect of the electron-electron interaction, we vary the Hubbard parameter $U$ but fix the Hund's coupling to \mbox{$J=0.65$\,eV}.
The local occupations and the spectral weight at the Fermi level are calculated from the imaginary time Green's function as \mbox{$n = -G(\beta)$} and \mbox{$\bar{A}(0) = - \tfrac{\beta}{\pi} \textrm{Tr}[G(\beta/2)]$}, respectively.
All results presented here are obtained from ``one-shot'' calculations, i.e., neglecting charge self-consistency.

\section{Results}
\label{sec:results}

\subsection{Octahedral rotations}
\label{subsec:structure}

We first analyze how the distortion of the octahedral network evolves within the different superlattices.
As already mentioned, bulk \lvo{} exhibits octahedral tilts and rotations, while bulk \svo{} crystallizes in a perfect cubic structure.
In Ref.~\cite{Luders_et_al:2014} it was observed that for a constant number of \lvo{} layers ($i=6$), the octahedral tilt and rotation pattern in \lvo{} remains bulk-like for a small number of \svo{} layers of up to $j<5$.
However, for more \svo{} layers, the experimental signal indicated complete suppression of tilts and rotations in \lvo{}.
In the following, we show that this trend cannot be observed in our calculations.
Instead, we find that for all considered multilayers the tilt angles show a continuous transition across the interface over a length scale of 4 unit cells, whereas the rotation angles are nearly unaffected by the multilayer geometry and change abruptly at the interface.

\begin{figure}
    \centering
        \includegraphics[width=\linewidth]{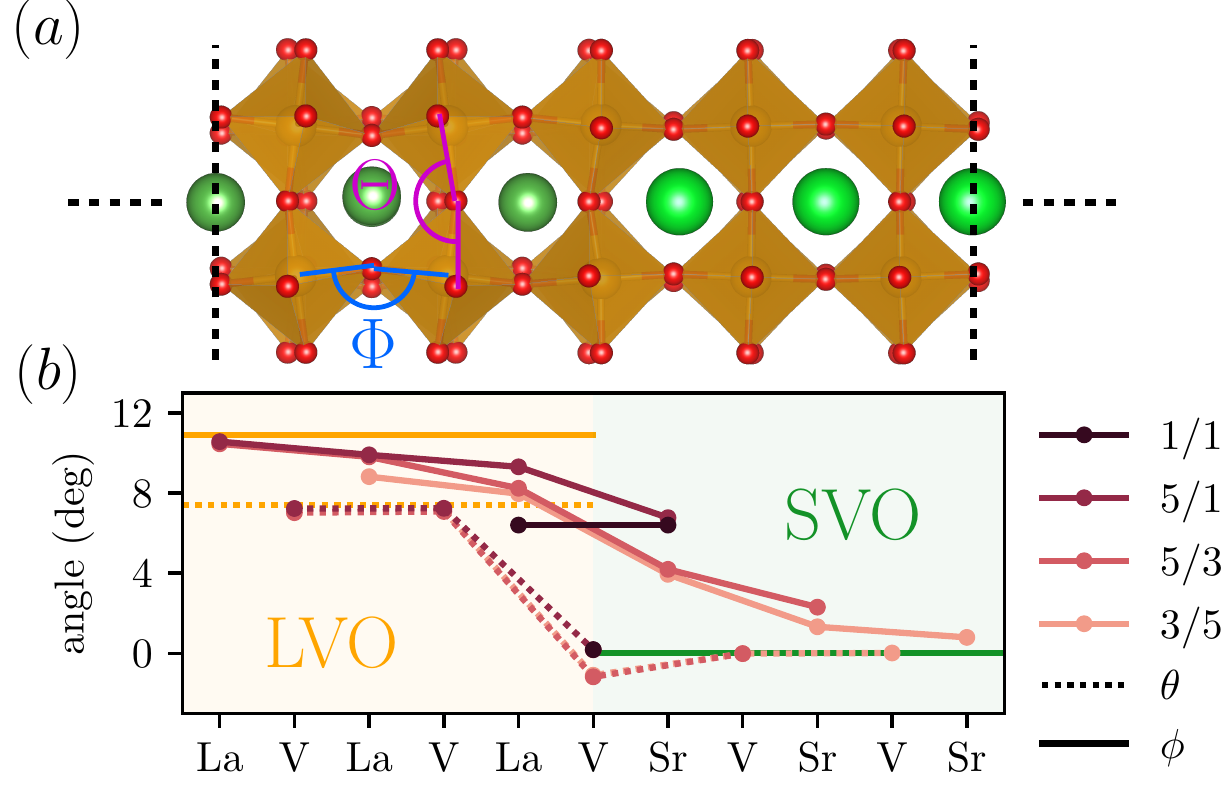}
    \caption{%
    (a) Unit cell of a 3/3 multilayer. La (Sr) atoms are shown in dark (bright) green, V in orange and O in red.
    From the TM-O-TM and O-O-O bond angles $\Phi$ and $\Theta$, we define the octahedral tilt and rotation angles as $\phi=(\pi-\Phi)/2$ and $\theta=(\pi/2-\Theta)/2$, illustrated by the blue and pink lines, respectively.
    (b) Evolution of octahedral tilts and rotations in the \lvo{}/\svo{} multilayers along the $z$ direction, represented by solid and dashed lines, respectively.
    The corresponding strained bulk values are shown in orange for \lvo{} and green for \svo{}.
    }
    \label{fig:ang_LS}
\end{figure}

\Cref{fig:ang_LS}(b) shows the layer-dependent tilt and rotation angles for a selected and representative number of superlattices.
As a reference, the corresponding bulk values for \lvo{} and \svo{} are indicated by the orange and green lines, respectively.
We first discuss the evolution of the in-plane octahedral rotation angles $\theta$, shown as dotted lines.
For all multilayers, the rotation angles show a sharp transition from a finite value in \lvo{}, essentially equal to the corresponding bulk value, to zero in \svo{}.
An interesting observation is the absence of rotations in the interfacial VO$_6$ layer, suggesting that on the structural level the influence of the Sr environment dominates over that of La.
In contrast, the tilt angles $\phi$ (solid lines) show a more continuous monotonic gradual reduction from the finite bulk value of \lvo{} to zero inside the \svo{} layers.
This transition occurs on a length scale of around 4 layers.
Interestingly, the 1/1 superlattice essentially exhibits an $a^-a^-c^0$ tilt system (in Glazer notation).

In comparison to experiment, our results do not provide evidence for complete suppression of tilts for a larger number of \svo{} layers.
It should be noted, however, that the growth direction in Ref.~\cite{Luders_et_al:2014} was oriented along the orthorhombic $[110]$ direction and thus differs from the setup used here.
As reported in Ref.~\cite{Sclauzero/Ederer:2015}, the out-of-plane tilts in \lvo{} are suppressed for this growth direction for a compressive strain above \SI{-3}{\%}.
Although in the experiment the strain on \lvo{} is only about \SIrange[range-phrase={ to }]{-0.5}{-1}{\%}, other structural reconstruction mechanism such as cation intermixing or off-stoichiometry may have contributed to a suppression of octahedral tilts and rotations.

\subsection{Electronic structure}
\label{subsec:ct}

\begin{figure}
    \centering
    \includegraphics[width=\linewidth]{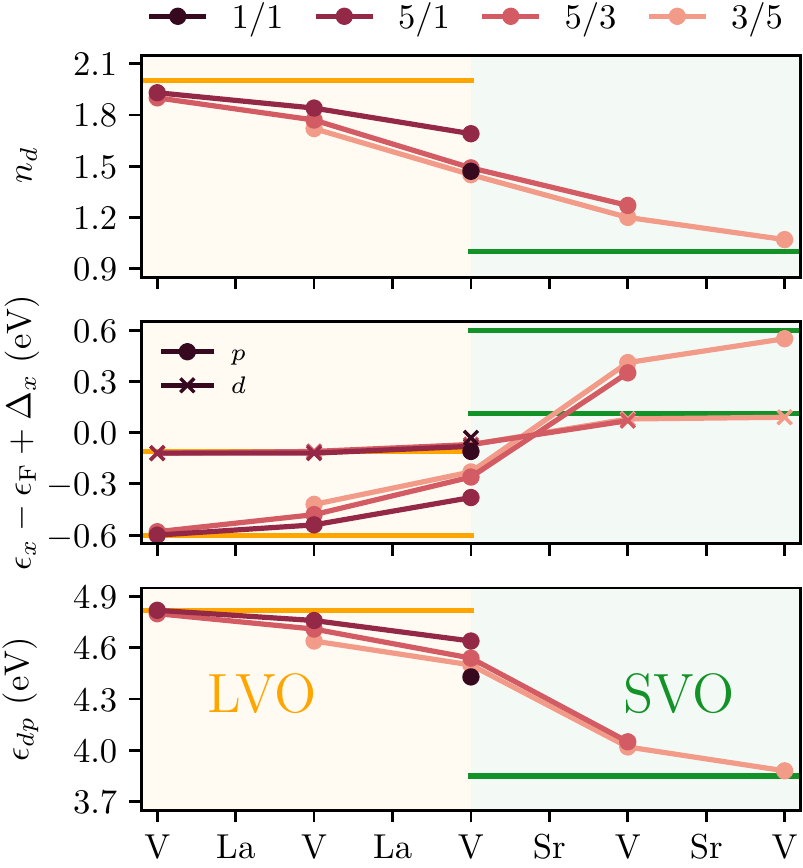}
    \caption{%
    Evolution of the changes in $d$ occupation (top), the O-$p$, and TM-$d$ (only $t_{2g}$) energies with respect to $\epsilon_\textrm{F} + \Delta_x$ ($x=p,d$), where $\epsilon_\textrm{F}$ is the Fermi level and $\Delta_p = \siev{4.88}$ and $\Delta_d = \siev{0.54}$ are arbitrary shifts introduced solely for better visualization on the same energy scale (middle), and the energy differences $\epsilon_{dp}$ (bottom) as a function of the TM site along the $z$ direction of the multilayers.
    }
    \label{fig:ct_LS}
\end{figure}

Next, we discuss the interface-related changes in the electronic states on the basis of the \ac{DFT} results.
\Cref{fig:ct_LS} shows the layer-dependent \ttg{} occupations, the charge transfer energy, $\epsilon_{dp} = \epsilon_d - \epsilon_p$, and the corresponding individual orbital energy levels for various multilayers with different thicknesses.
Here, the occupations are computed from the layer-resolved density of states by integrating all atomic contributions per VO$_2$ layer over the energy region of the \ttg{} bands up to the Fermi level.
To extract the energy levels corresponding to atomic-like $p$ and $d$ orbitals we construct a larger set of Wannier functions that includes the vanadium \ttg{}-derived states as well as the oxygen $p$ states.
The label $d$ therefore refers to \ttg{} states only, while the energies of the $\epsilon_p$ are taken as the averaged Wannier levels of all oxygen $p$ sites in the VO$_2$ layers.

We first focus on the evolution of the vanadium $3d$ occupations in the upper panel.
Within the interfacial layer, the coordination of the vanadium is associated equally with \lvo{} or \svo{} on either side.
As can be seen, this leads to a mixed oxidation state of $+3.5$, or an occupation of \SI{1.5}{\elc}, of the interfacial V cation, corresponding to the average over the two bulk compounds.
The only exception occurs for a single dopant layer of SrO embedded in \lvo{} (i.e., the 5/1 multilayer in \Cref{fig:ct_LS}), where the occupation in the interfacial layer is further increased towards the \dtwo{} valence of \lvo{}.
Away from the interface, the occupations quickly converge to those in the corresponding bulk systems within about two layers, i.e., on a length scale very similar to that of the octahedral tilts discussed in Sec.~\ref{subsec:structure}.
For all multilayers, the charge is consistently transferred from \lvo{} to \svo{}, i.e., from the transition metal with a \dtwo{} valence to that with a \done{} valence electron configuration, in accordance to earlier \dftu{} studies~\cite{Park/Kumar/Rabe:2017}.

The direction of the charge transfer can be understood in terms of the occupancy difference of the $d$ levels, as discussed in Ref.~\cite{Chen/Millis:2017}. 
The average V oxidation state of $+3.5$ in the interfacial layer smooths the occupancy discontinuity between the nominal bulk oxidation states of the V sites, $+3$ and $+4$ for \lvo{} and \svo{}, respectively.
Furthermore, inspecting the evolution of the charge transfer energy, $\epsilon_{dp}$ (bottom panel of \Cref{fig:ct_LS}), which can be related to the electronegativity of the transition metal cations relative to the oxygen ligands, we find that the evolution of $\epsilon_{dp}$ across the interface essentially mirrors the changes in the occupancy.
Away from the interface, the difference in $\epsilon_{dp}$ becomes identical to that of the two bulk compounds (shown as orange and green lines), which is of the order of \siev{1.0}. The lower $\epsilon_{dp}$ in \svo{} relative to \lvo{} indicates a higher electronegativity of the V$^{4+}$ cation in \svo{}, consistent with the observed charge transfer.

The middle panel of \Cref{fig:ct_LS} illustrates that the variation of $\epsilon_{dp}$ is mainly due to the difference in the oxygen $p$ levels, $\Delta\epsilon_p\sim\siev{1.2}$, between the two bulk values, while the difference of the V $d$ levels reduces $\epsilon_{dp}$ by approximately \siev{0.2}.
We find a gradual deviation of oxygen $p$ levels from the corresponding bulk values by up to \siev{0.3} on either side of the interface, smoothing the transition of oxygen $p$ levels across the interface.
This is in line with the ``oxygen continuity condition'' proposed in Ref.~\onlinecite{Zhong/Hansmann:2017}, stating that the $p$ states of the oxygen network tend to align continuously across the interface.

\subsection{\ac{DMFT} results}
\label{subsec:dmftsvo}

We now discuss the \ac{DMFT} results for selected multilayers.
\begin{figure}
    \centering
    \includegraphics[width=\linewidth]{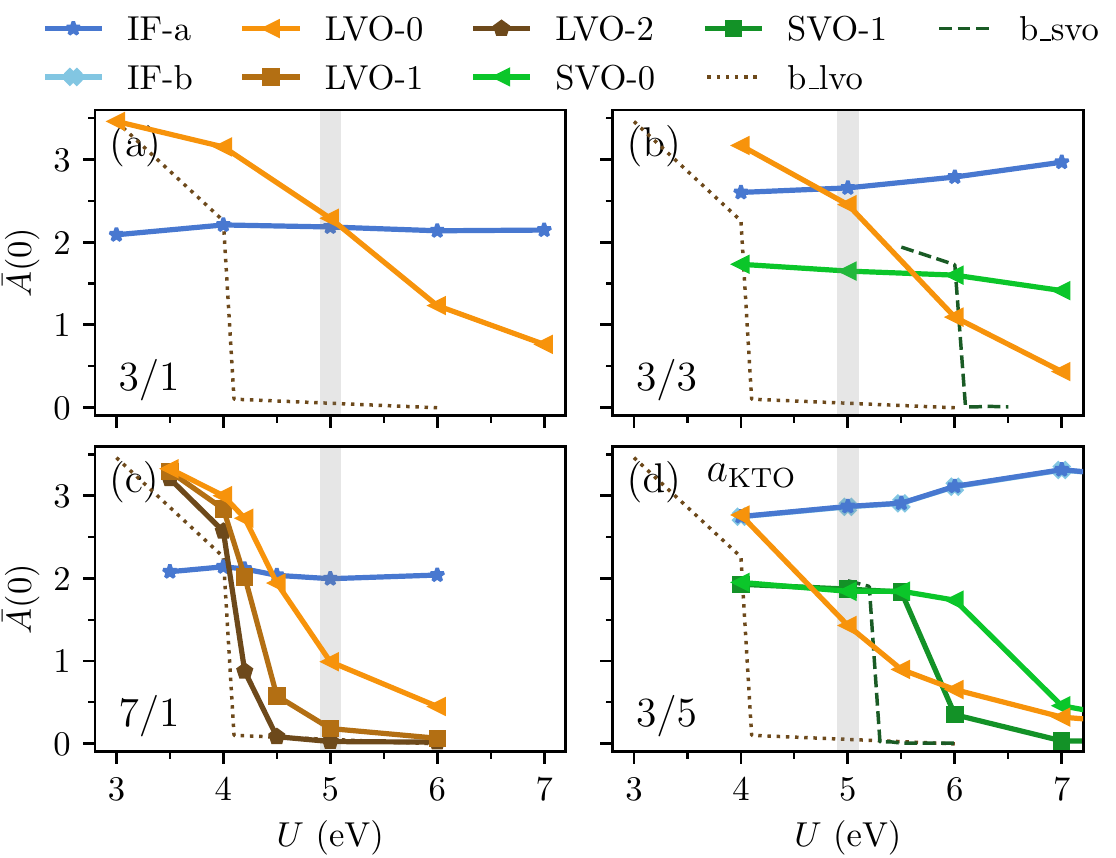}
    \caption{%
    Averaged spectral weight at the Fermi level, $\bar{A}(0)$, as a function of the interaction parameter $U$ for different VO$_2$ layers in different multilayer geometries.
    Orange lines and symbols refer to \lvo{}, while green ones represent \svo{} layers, with $n$ in $A$VO-$n$ indicating the distance from the interface (and $A$=L or S for \lvo{} and \svo{} layers, respectively). Interface layers are denoted as IF (see main text for further details).
    The reference data for (unstrained) bulk \lvo{} (b\_lvo) and \svo{} (b\_svo) are indicated by black dotted and dashed lines, respectively.
    }
    \label{fig:dmft_LS}
\end{figure}
%
\begin{figure}
    \centering
    \includegraphics[width=\linewidth]{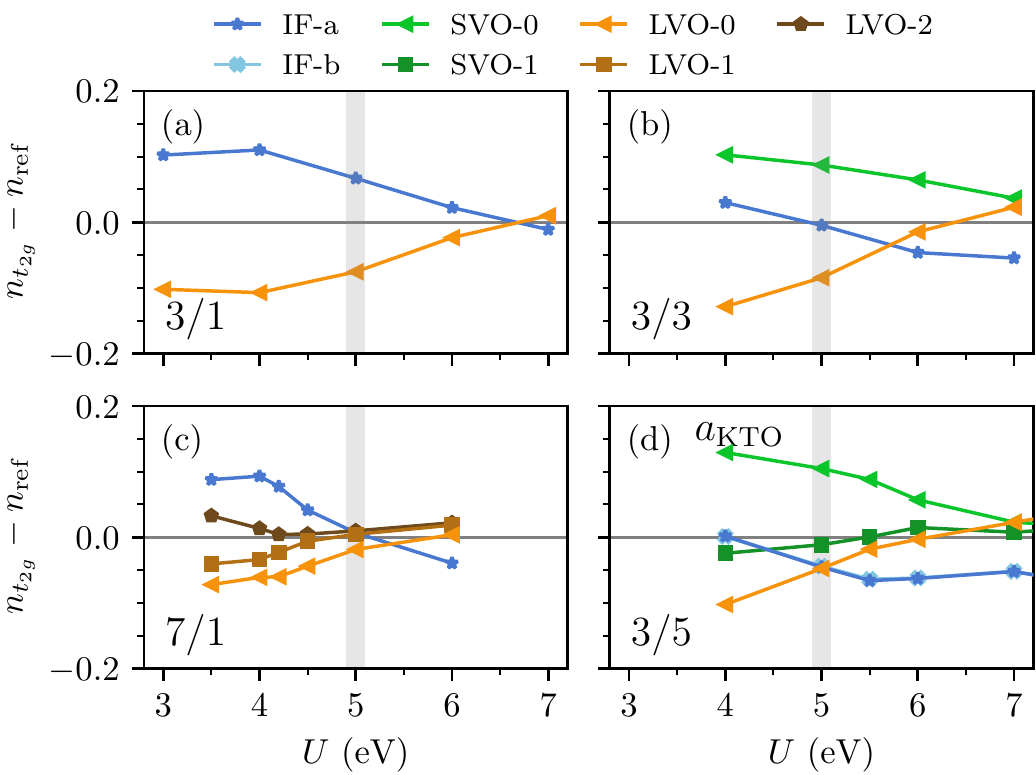}
    \caption{%
    Total \ttg{} occupation of symmetry-inequivalent VO$_2$ layers in different multilayers as a function of the interaction parameter $U$. For better visualization, all occupations are plotted relative to ``reference occupations'' $n_{\textrm{ref}}$, which are taken to be equal to 2 and 1 electrons for \lvo{} and \svo{} layers, respectively, and equal to 1.5 for the interface layers (IF).
    The notation used to denote different layers is the same as in \Cref{fig:dmft_LS}.
    }
    \label{fig:dmft_occ}
\end{figure}
The first question we address concerns the thickness dependence of electronic reconstruction within the \lvo{} layers.
For this, we analyze the layer-dependent spectral weight at the Fermi level, $\bar{A}(0)$, as well as the layer-dependent \ttg{} occupations as function of the interaction strength, represented by the Hubbard $U$ parameter.

In the 3/1 case, shown in \Cref{fig:dmft_LS}(a), the multilayer consists of four layers, with two symmetry-equivalent \lvo{} layers with bulk coordination, denoted as LVO-0 and shown in orange, and two symmetry-equivalent interfacial VO$_2$ layers, shown in blue and labeled as IF-$a$.
It can be seen that, for increasing $U$, the spectral weight at the Fermi level remains constant in the interfacial layer, while it is strongly decreased within the more bulk-like \lvo{} layer.
Simultaneously, the local electron-electron interaction establishes ``nominal'' occupations of approximately 2 and 1.5 in the LVO-0 and IF-$a$ layers, respectively, for $U \approx \siev{7}$ (see \Cref{fig:dmft_occ}(a)).
Thus, for a typical interaction strength of  $U\approx\siev{5}$~\cite{Pavarini_et_al:2004,DeRaychaudhury/Pavarini/Andersen:2007,Sclauzero/Ederer:2015}, all layers are metallic, and remain so up to at least $U=\siev{7}$.

Increasing the \lvo{} layer thickness, as shown for the 7/1 multilayer in \Cref{fig:dmft_LS}(c), reveals that already the second layer away from the interface, denoted as LVO-1, has only minimal remaining spectral weight at the Fermi level for $U=\siev{5}$, while the innermost layer, LVO-2, is essentially already bulk-like.
Interestingly, even the first layer, LVO-0, is evidently more bulk-like (with strongly reduced spectral weight above $\sim\siev{4}$) compared to the corresponding layer in the 3/1 multilayer.
This is a result of the transferred charge being redistributed over a larger number of \lvo{} layers, which are thus closer to the nominal occupancy, as shown in \Cref{fig:dmft_occ}(c).
As for the 3/1 case, the metallic character of the interfacial VO$_2$ layer, IF-$a$, remains unaffected by changes in the interaction parameter.
The metallicity of the interfacial layer seems to prevent the occupations to adopt exact integer values even for large $U$, in contrast to what has been observed, for example, from a similar analysis of charge transfer between two Mott insulators (see Fig. 4 in Ref.~\cite{Beck/Ederer:2019}).
Interestingly, the small deviations from exact integer filling do not seem to disturb the Mott-insulating state within the \lvo{} layers, as can be seen from the vanishing spectral weight of LVO-2 in \Cref{fig:dmft_LS}(c) for large $U$, despite of its non-integer occupation.

From this we conclude that, despite the doping, bulk-like behavior is restored in \lvo{} within \numrange{2}{3} layers away from the interface. 
Thus, in contrast to Ref.~\cite{Sheets/Mercey/Prellier:2007}, but consistent with the later reports in Ref.~\cite{Luders_et_al:2009}, our results show no indications of a filling-controlled insulator-to-metal transition for $m$/1 superlattices with $m \leq 4$.
Even though the \lvo{} layers further away from the interface become insulating, the 7/1 multilayer is still expected to have a conducting channel at the interface, which would lead to metallic in-plane conductivity. On the other hand, out-of-plane transport measurements would indeed be expected to exhibit insulating behavior for sufficiently thick \lvo{} layers.
Furthermore, comparing to the solid solution, the 7/1 multilayer would correspond to a doping concentration of $x=0.125$ and would thus be expected to be a paramagnetic insulator at room temperature\cite{Miyasaka_et_al:2003}.
Our result thus further demonstrates the distinct behavior of heterostructures compared to solid solutions, thanks to the spatially coherent arrangement of the chemical constituents.

Next, we consider the effect of increasing the number of \svo{} layers, using the 3/3 multilayer as an example, shown in \Cref{fig:dmft_LS}(b).
While the behavior of the \lvo{} layer remains unchanged compared to the 3/1 case, the additional \svo{} layer seems to enhance the metallic regime at the interface and within the \svo{} layer.
This is perhaps not surprising considering that \svo{} is a correlated metal with a finite quasiparticle weight for $U=\siev{5}$.
However, as indicated by the dashed dark green line in \Cref{fig:dmft_LS} (labeled b\_svo), bulk \svo{} undergoes an \ac{MIT} at $\umit{}\approx\siev{6}$, while the single \svo{} layer in the 3/3 multilayer shows no such transition.
This indicates that the charge reconstruction at the interface, extending into the adjacent \lvo{} and \svo{} layers (see \Cref{fig:dmft_occ}(b)), stabilizes the metallicity.
Our results further confirm the absence of a dimensionality-induced metal-insulator transition reported for ultra-thin \svo{} films grown on \sto{}~\cite{Yoshimatsu_et_al:2010,Zhong_et_al:2015} when buried instead in \lvo{} layers~\cite{Li_et_al:2015}.

At this point, we briefly summarize that our \ac{DMFT} results indicate that the interfacial layer is persistently metallic for all multilayers.
While on the \lvo{} side the Mott-insulating state and approximately nominal occupations are recovered on a length scale of \numrange{2}{3} unit cells, the metallicity on the \svo{} side is enhanced compared to bulk \svo{}.
The two questions we want to address next are (a) whether the thickness of the metallic layer can be tuned and (b) under what circumstances the two interfaces can be brought into a charge/interfacial ordered state with integer occupations that could promote a Mott-insulating state across all layers. We first focus on question (a), while (b) is discussed in \Cref{subsec:strain}.

To address the first question, we show in \Cref{fig:dmft_LS}(d) a calculation for a strained superlattice of three layers of \lvo{} and five layers of \svo{} (3/5 multilayer), where the in-plane pseudo-tetragonal lattice parameter is fixed to the bulk lattice constant of KTaO$_3$.
This choice is based on the fact that KTaO$_3$, if chosen as a substrate, would impose a sufficiently large lattice mismatch with respect to \svo{} to result in an \ac{MIT}~\cite{Sclauzero/Dymkowski/Ederer:2016}.
For the experimental lattice constant of bulk KTaO$_3$~\cite{Wyckoff:1935}, \mbox{$a_{\mathrm{KTO}}= \SI{3.981}{\angstrom}$}, the tensile strain amounts to \SI{2.2}{\percent} and \SI{3.2}{\percent} in \lvo{} and \svo{}, respectively.
While this is not expected to lead to noticeable changes within \lvo{}~\cite{Sclauzero/Dymkowski/Ederer:2016}, a comparable magnitude of tensile strain shifts \svo{} towards the insulating regime~\cite{Sclauzero/Dymkowski/Ederer:2016}, indicated by the calculated decrease of \umit{} from \siev{6} to slightly above \siev{5} in the strained bulk reference shown in \Cref{fig:dmft_LS}(d).

As seen by the sudden decrease of spectral weight for $U\geq\siev{5.5}$ in the SVO-1 layer, and at somewhat higher $U$ also in the SVO-0 layer, the tensile strain indeed supports formation of the Mott-insulating phase in \svo{} also in the multilayer, similar to what has been found for the bulk case~\cite{Sclauzero/Dymkowski/Ederer:2016}.
On the other hand, even in this case the interfacial layer, IF-$a$, remains persistently metallic, with a high spectral weight around the Fermi level, even for large $U$ values.
As expected, the \lvo{} layer is only marginally affected by the tensile strain.
The occupations in \Cref{fig:dmft_occ}(d) again show a tendency towards integer bulk occupations under increasing $U$, without actually reaching exact integer values even for large interaction parameter.
These results indicate that for large tensile strain it might be possible to obtain insulating solutions on both sides of the interface, while the interface itself remains metallic, with the resulting metallicity being strongly two-dimensional, i.e., localized within a single VO$_2$ layer.

\subsection{Strain-modulated multilayers}
\label{subsec:strain}

Ref.~\cite{Tan_et_al:2013} reported an atomic scale oxidation map of (\lvo{})$_{6}$/(\svo{})$_3$ superlattices via STEM-EELS measurements, revealing an electronic asymmetry in the valence configuration between the ``upper'' and ``lower'' \lvo/\svo{} interfaces (relative to the growth direction), despite the interfaces being chemically symmetric.
The oxidation states of the interfacial VO$_2$ layers were found to range between the expected $+3$ in \lvo{} and $+4$ in \svo{}, but were not identical at the two interfaces.
This was associated with the observation of a local strain modulation along the $z$-direction of the superlattice.
Consistent with the expected epitaxial strain, the cell parameters along the $z$ direction of the films were found to vary by about $+1.5$\,\% within \lvo{} and by \SI{-3}{\%} within \svo{} with respect to the in-plane lattice constant fixed to the \sto{} substrate.
Interestingly, the modulation of the lattice parameter along the $z$ direction showed an asymmetric profile, with a more abrupt change at one interface, and a  more gradual change at the other. This was interpreted as a result of strain relaxation during the growth process~\cite{Matthews/Blakeslee:1974}.

Here, we explore the possibility of obtaining such an interfacial charge imbalance by imposing different strain modulations along the $z$-axis.
As mentioned in the previous section, a strong interfacial asymmetry, leading to integer occupations of the interfacial V cations, could in principle allow for a global Mott insulating state in the multilayer.
We note that an in-plane charge order within each interface, as explored with \dftu{} in Refs.~\cite{Park/Kumar/Rabe:2017,Dai_et_al:2018}, provides an alternative route to a possible insulating state at the interface, which we do not pursue here.

\begin{figure}
    \centering
    \includegraphics[width=\linewidth]{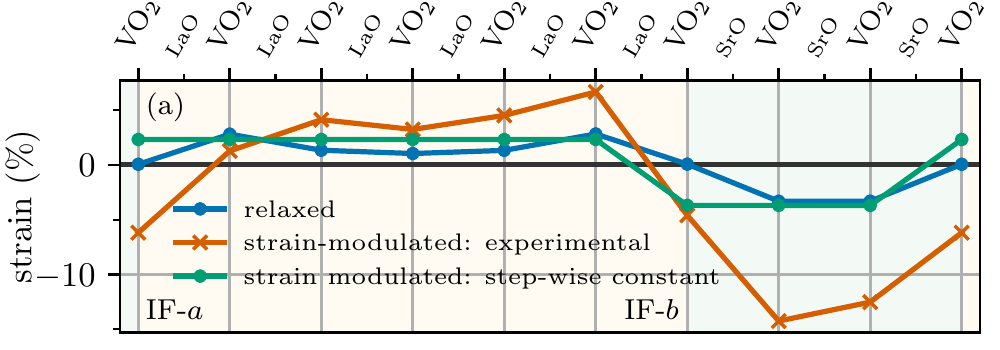}\\
    \includegraphics[width=\linewidth]{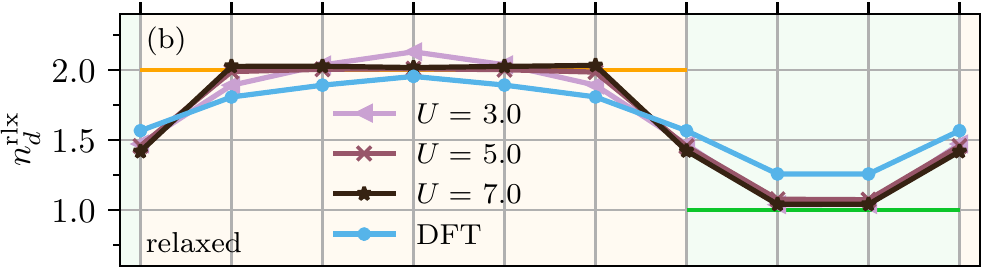}\\
    \includegraphics[width=\linewidth]{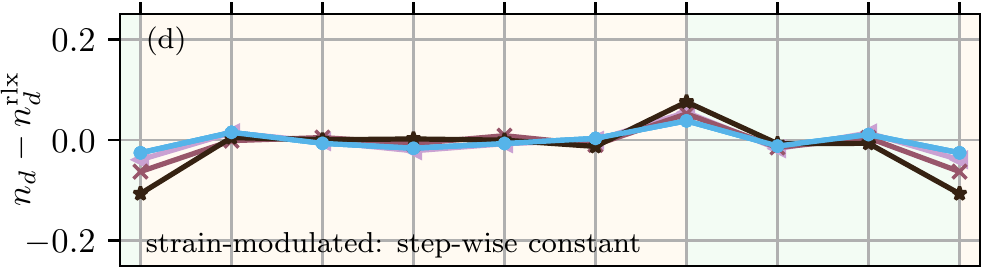}\\
    \includegraphics[width=\linewidth]{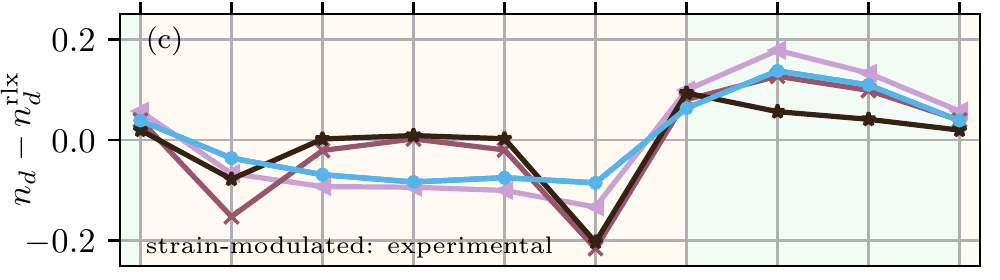}
    \caption{%
    (a) Strain profiles along the $z$-axis of the 6/3 multilayer with respect to the in-plane lattice constant. 
    (b) Layer-resolved V-$d$ site occupancies along the $z$-axis of the relaxed structure for the non-interacting case (light blue) and for various interaction parameters.
    The reference values for bulk \lvo{} and \svo{} are shown as orange and green lines, respectively.
    (c-d) Layer-resolved difference in occupancy of the strain-modulated structures with respect to the relaxed structure in (b).
    }
    \label{fig:LS_set}
\end{figure}

Analogous to Ref.~\cite{Tan_et_al:2013}, we use a 6/3 multilayer, for which we consider  three different strain profiles along the $z$-axis, as shown in \Cref{fig:LS_set}(a): i) the fully relaxed case, ii) the experimental strain profile enhanced by a factor of five, and iii) a case with constant strain levels in the two components but asymmetric interfaces. 
Here, the strain in each VO$_2$ layer is defined as the distance between $A$ (La/Sr) sites in the adjacent $A$O layers, relative to the fixed in-plane lattice constant. 

The experimental strain profile is extracted from Fig. 1(b) of Ref.~\cite{Tan_et_al:2013}, averaged over three superlattice periods.
The $A$ site positions as well as the length of the simulation cell along $z$ are then adjusted accordingly and kept fixed, while all other internal coordinates are relaxed.
To potentially obtain a stronger effect of the strain modulation, the magnitude of the experimental strain distribution is enhanced by a factor of five.
For the simplified case with step-wise constant strain modulation (green line in \Cref{fig:LS_set}(a)), we keep the length of the simulation cell equal to the relaxed case, but use a fixed spacing between all $A$ sites within six ``nominal \lvo{} layers'' (including IF-$a$) and analogous for the remaining three ``nominal \svo{} layers'' (including IF-$b$). Thus, the two interfacial VO$_2$ layers, IF-$a$ and IF-$b$, are structurally associated with \lvo{} and \svo{}, respectively.
The relative strain levels in the two parts of the slab are chosen to be consistent with experiment~\cite{Tan_et_al:2013}, which results in a positive strain of \SI{2.3}{\%} for \lvo{} and a negative strain of \SI{-3.7}{\%} for \svo{}. 
Also in this case, all internal coordinates except for the $z$-coordinates of the $A$ sites are relaxed.

We note that, in contrast to the cases with odd numbers of LaO and SrO layers considered in the previous sections, the two interfaces within the 6/3 slab, termed IF-$a$ and IF-$b$, are no longer related by a specific symmetry operation, since the potential mirror symmetry relative to the central VO$_2$ layers is broken by the presence of the octahedral tilts.

We first discuss the relaxed case, without an imposed strain modulation.
The corresponding layer-dependent V-$d$ occupations, shown in \Cref{fig:LS_set}(b), exhibit the same behavior as the cases already discussed in \Cref{subsec:dmftsvo}.
The occupations of the layers away from the interface quickly converge towards the corresponding bulk values, while the interfaces remain fractionally occupied, corresponding to an intermediate oxidation state of $+3.5$, for all considered interaction parameters. Thus, even though this is no longer strictly enforced by symmetry, IF-$a$ and IF-$b$ remain electronically equivalent.

\Cref{fig:LS_set}(c) shows the layer-dependent change in V-$d$ occupation relative to the relaxed case for the enhanced experimental strain profile.
The large average strain levels in the different parts of the multilayer lead to an enhanced overall charge transfer from \lvo{} to \svo{}.
Thereby, for low values of $U$, we find that the changes in occupation (relative to the relaxed case) seem to approximately mirror the applied strain profile, but with opposite sign. Thus, a local compression (elongation) along $z$ leads to an accumulation (depletion) of charge in the corresponding layer.
For higher $U$ values, the innermost layers in \lvo{} (and for $U=\SI{7}{eV}$ also in \svo{}) are pushed back towards an integer occupancy, which enhances the charge imbalance in the sub-interface \lvo{} layers. Even though there is a small difference in occupation between the two interface layers, this difference remains very small for all values of $U$. Furthermore, both interfaces exhibit a positive charge accumulation relative to the relaxed case. Thus, the resulting charge distribution appears to be rather complex, with competing effects of local strain and interaction, and does not straightforwardly compare to the experimental observations.

In order to get a more systematic insight, we therefore consider the case with a simple step-wise constant strain modulation shown in \Cref{fig:LS_set}(d).
In this case, the occupation change is very small ($\leq 0.02$) in all inner layers, both in the \lvo{} part and in the \svo{} part.
The only noticeable effect is seen in the two interface layers, where IF-$a$ (IF-$b$) loses (gains) charge, resulting in a sizeable charge difference of up to 0.11 per interface, or 0.22 in terms of charge asymmetry between the interfaces (depending on $U$).
However, even for $U=\siev{7}$ the resulting charges are far from reaching integer values.
Thus, the simple step-wise variation of strain leads only to a marginal charge imbalance, which does not seem to support a fully polarized state of V$^{4+}$ to V$^{3+}$ at the respective interfaces that could result in an overall Mott-insulating system.
Furthermore, we find that the sign of the occupation change exhibits the opposite trend that one might expect based on the incentive imposed on the structure.
That is, IF-$a$, which is structurally associated with \lvo{}, tends to a $+4$ oxidation state, while IF-$b$ tends to $+3$.

In conclusion, our results demonstrate that an inhomogeneous strain modulation along the $z$-axis of the superlattice can in principle lead to a charge imbalance at the interfaces. However, the resulting effect seems rather weak and is unlikely to be sufficient to obtain integer valences at the interfaces, which would be required to obtain a Mott-insulating state for sufficiently strong interaction parameters.
This further confirms the results presented in \Cref{subsec:dmftsvo} that the interface metallicity is very robust.
Furthermore, imposing the experimentally measured strain profile in our calculations, even with strongly enhanced amplitude, does not reproduce the experimentally observed valence profile even qualitatively. This indicates that factors other than pure strain, such as, e.g., more complex structural relaxations, cation intermixing, the specific substrate orientation, defects, etc., are likely to be relevant.

\section{Conclusions}
\label{sec:sum}

We investigated the structural and electronic reconstruction of multilayers consisting of the Mott insulator \lvo{} and thin layers of the correlated metal \svo{}.
For the $[001]$ substrate orientation studied here, the octahedral tilts present in \lvo{} penetrate slightly into the first few \svo{} layers, while the octahedral rotations around $[001]$ drop to zero abruptly at the interface. 
In addition, there is electronic charge transfer from \lvo{} towards \svo{} across the interface, with the V-cations in the interfacial layers assuming an intermediate oxidation state of $+3.5$, except for the case of a single SrO layer embedded within \lvo{}, where the corresponding $d$-orbital occupation becomes larger than 1.5$e^-$.
Thus, the insertion of even a single layer of SrO into \lvo{} leads to hole doping of the Mott insulator and a metallic layer that persists even for large interaction parameters, even though a large $U$ tends to drive the occupation towards integer values.
Both on the structural and electronic level, we find that the length scale of reconstruction is of the order of two unit cells, which is consistent with earlier studies of similar systems~\cite{Okamoto/Millis:2004, Tan_et_al:2013}.

While a gap in the local spectral function is recovered in \lvo{} layers away from the interface, the metallicity in \svo{} is stabilized.
This indeed suggests a picture of geometrically confined doping~\cite{Luders_et_al:2009} in buried layers of \svo{}, in contrast to the dimensionality-induced metal-insulator transition in ultra-thin films of \sto{}~\cite{Yoshimatsu_et_al:2010}.
Our results therefore highlights the importance of the spatial arrangement of the atoms in multilayer geometries compared to simple solid solutions with identical overall composition, which opens new possibilities for designing artificial functional materials.

We further demonstrated that an inhomogeneous strain modulation applied along the growth direction introduces additional charge reconstruction that can create electronically asymmetric interfaces. 
However, the effect is relatively weak and is therefore not sufficient to result in a Mott insulating state with integer occupations across all layers.

\appendix*

\begin{acknowledgments}

This work was supported by ETH Zurich and the Swiss National Science Foundation through NCCR-MARVEL.
Calculations were performed on the cluster ``Piz Daint'' hosted by the Swiss National Supercomputing Centre.
The Flatiron Institute is a division of the Simons Foundation.

\end{acknowledgments}

\bibliography{main}

\end{document}